\documentclass[twocolumn, prl, superscriptaddress,notitlepage]{revtex4-2}
\usepackage{graphicx}
\usepackage{dcolumn}
\usepackage{amsthm}  
\usepackage{amsmath} 
\usepackage{comment}
\usepackage{bm}
\usepackage[usenames,dvipsnames]{color}
\usepackage[most]{tcolorbox}
\usepackage{multirow}
\usepackage{gensymb}
\usepackage{amssymb}
\usepackage{amsmath}
\usepackage{xcolor}
\usepackage{algorithm}
\usepackage{algorithmic}
\usepackage[normalem]{ulem}
\usepackage{CJK}
\usepackage{comment}
\usepackage{amsfonts}
\usepackage[colorlinks, linkcolor=blue,anchorcolor=blue,citecolor=blue,urlcolor=blue]{hyperref}
\usepackage{amssymb}
\usepackage{pifont}
\usepackage{physics}
\usepackage{natbib}
\usepackage{xcolor}

\usepackage{color,soul}

\begin{document}

\title{Optimal scheme for distributed quantum metrology}

\author{Zhiyao Hu }\email[]{zhiyaohu.phys@gmail.com}

\affiliation{Pritzker School of Molecular Engineering, The University of Chicago, Chicago, Illinois, USA}

\author{Allen Zang}
\affiliation{Pritzker School of Molecular Engineering, The University of Chicago, Chicago, Illinois, USA}

\author{Jianwei Wang}
\affiliation{Department of Physics, University of Chicago, Chicago, Illinois, USA}

\author{Tian Zhong}\email[]{tzh@uchicago.edu}
\affiliation{Pritzker School of Molecular Engineering, The University of Chicago, Chicago, Illinois, USA}

\author{Haidong Yuan}\email[]{hdyuan@mae.cuhk.edu.hk}
\affiliation{
   Department of Mechanical and Automation Engineering, The Chinese University of Hong Kong, Shatin, Hong Kong}

\author{Liang Jiang}\email[]{liang.jiang@uchicago.edu}
\affiliation{Pritzker School of Molecular Engineering, The University of Chicago, Chicago, Illinois, USA}

\author{Zain H. Saleem}\email[]{zsaleem@anl.gov}
\affiliation{Mathematics and Computer Science Division, Argonne National Laboratory, Lemont, IL, USA}

\begin{abstract}

Optimal strategies for local quantum metrology---including the preparation of optimal probe states, implementation of optimal control and measurement strategies, are well established. However, for distributed quantum metrology, where the goal is to estimate global properties of multiple spatially distributed parameters, the optimal scheme---particularly the role of optimal control---remains poorly understood. In this work, we address this challenge by developing optimal schemes for distributed quantum metrology that characterize the ultimate precision limits in distributed systems. We derive the optimal probe state, optimal control protocols, and measurement strategies in estimating a linear combination of $N$ independent unknown parameters coupled to $d$ networked sensors. Crucially, we prove that the optimal control operations can be implemented locally on each sensor, eliminating the need for non-local control operations across distant nodes. This result significantly reduces the complexity of implementing optimal strategies in distributed quantum metrology. To demonstrate the power of our framework, we apply it to several key scenarios.

\end{abstract}
\maketitle

\paragraph{Introduction}---
Distributed quantum metrology (DQM), which seeks to measure global properties of multiple unknown parameters encoded across a network of spatially distributed sensors \cite{dqszhang2021distributed},
has gained increasing attention \cite{dqszhuang2018distributed,dqsguo2020distributed,dqsliu2021distributed,dqsmalia2022distributed} due to its wide applications, such as quantum radar \cite{qrmaccone2020quantum,qrassouly2023quantum} and global clock synchronization \cite{qcgiovannetti2001quantum,qckomar2014quantum}. 
Recent studies have shown that global entanglement in the probe state can improve the precision for estimating a linear combination of multiple time-independent signals encoded across distributed nodes \cite{qn2018multiparameter, qneldredge2018optimal, qnge2018distributed, qnqian2019heisenberg, qnqian2021optimal, qnehrenberg2023minimum, qnpezze2024distributed,zang2024quantum,oh2022distributed,malia2022distributed,zhuang2020distributed,gessner2020multiparameter,guo2020distributed,xia2020demonstration,liu2021distributed,zhao2021field}. However, control operations—often critical for achieving ultimate precision—remain largely unexplored in DQM. While systematic methods for optimizing control operations exist in local estimation scenarios \cite{REgiovannetti2011advances, QECdur2014improved,  qmnaghiloo_achieving_2017, qmpezze2017optimal, qmpang_optimal_2017, controlyuan2017quantum, REdegen2017quantum, qmpoggiali2018optimal, controlcontrolliu2017control, QEClayden2019ancilla, QECgorecki2020optimal, REpolino2020photonic, qmyang_variational_2022, REliu2022optimal, REliu2024fully, QECchen2024quantum, hu2024control, isogawa2025entanglement,gessner2018sensitivity}, extending these approaches to DQM presents unique theoretical and practical challenges. The fundamental challenge stems from the locality constraint: just as distributed sensing favors local measurements, control operations should ideally be performed locally. However, this preference significantly complicates the optimization of distributed quantum sensing protocols.

In this work, we study the optimal strategies for estimating arbitrary linear combinations of parameters encoded across a network of quantum sensors. Our analysis establishes a bound on the maximum achievable quantum Fisher information (QFI) for distributed sensing and simultaneously identifies (1) an optimal entangled probe state, (2) an efficient local control protocol, and (3) an optimal local measurement strategy. Surprisingly, while global entanglement in the initial probe state is generally required, we demonstrate that subsequent control and measurement operations need only be local, proving that nonlocal interventions during sensing are unnecessary. We validate the practical utility of our framework through examples that can find applications in global clock synchronization, quantum radar, and distributed magnetic field sensing.

\begin{figure}[htbp]
\centering
\includegraphics[width=0.45\textwidth]{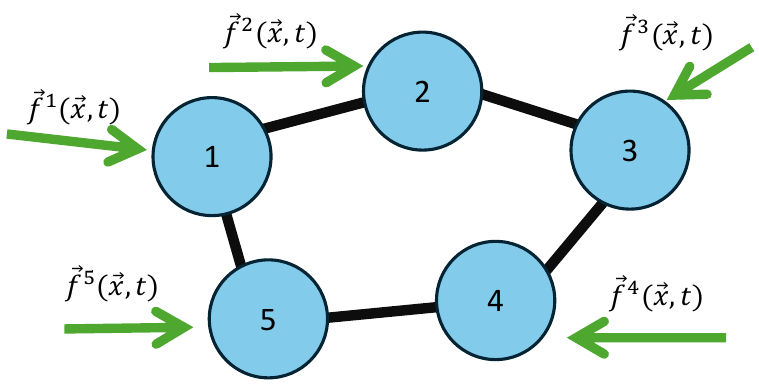}
\caption{A demonstration of networked quantum sensors. Here $\vec{f}(\vec{x},t)$ is the external field that processes the unknown parameters $\vec{x}$ that we hope to measure, and the coupling between the external field and the $i$-th node can be described as $H_{i}=\vec{f}^{i}(\vec{x},t)\cdot\vec{\sigma}^i$.
}
\label{fig1}
\end{figure}

\paragraph{Problem formulation}---
We consider a distributed quantum sensing system comprising $d$ sensor nodes, where each node $i$ consists of a qubit coupled to a general time-dependent vector field  $\Vec{f}^i(\Vec{x},t)=[f^{i}_x(\Vec{x},t),f^{i}_y(\Vec{x},t),f^{i}_z(\Vec{x},t)]$. The system's total Hamiltonian is given by $H_{\rm tot}=H_{\rm free}+H_C$, where $H_C$ represents externally tunable control operations and $H_{\rm free}$ describes the interaction between the sensors and the external field,
\begin{equation}
        H_{\rm free}=\sum_i^d \Vec{f}^i(\Vec{x},t)\cdot \Vec{\sigma}^i,
\end{equation}
here $\Vec{x}=(x_1, x_2,...,x_N)^T$ denotes the vector of unknown parameters, $\Vec{\sigma}^i$ denotes the Pauli matrices at node $i$. We are interested in characterizing the global properties of the parameters, specifically a linear combination \(\theta_1 = \vec{w}^T\vec{x}\) with $\vec{w}^T=(w_1,\cdots, w_N)$. 

The variance of any unbiased estimator of $\theta_1$ is lower bounded as
\begin{equation} 
\delta \hat{\theta}_1^2\geq \frac{\vec{w}^T\vec{w}}{\mu \vec{w}^TJ_{\vec{x}}(T)\vec{w}},
\end{equation}
which is known as the weak form of Cramer-Rao bound (CRB) \cite{gessner2018sensitivity,dqskim2024distributed}, here $\delta \hat{\theta}_1^2$ denotes the variance, $\mu$ denotes the number of repeated measurements. 
Defining the quantity $\vec{w}^T J{\vec{x}}(T) \vec{w}$ as the effective quantum Fisher information (QFI) $J_{\theta_1}(T)$, we are interested in identifying the maximal achievable value of $J_{\theta_1}(T)$.

Let $U_{\rm tot}(T)$ be the unitary operator generated by $H_{tot}$ at time $T$, the generator for $x_j$ is then given by $S_{x_j}(T)=i U_{\rm tot}^\dagger(T) \partial_{x_j} U_{\rm tot}(T)$. The entries of the QFIM $J_{\vec{x}}(T)$ can then be obtained as $J_{x_ix_j}(T)=4{\rm Re}(\langle S_{x_i}(T)S_{x_j}(T)\rangle-\langle S_{x_i}(T)\rangle\langle S_{x_j}(T)\rangle)$\cite{qmhelstrom1976sld,qcrbfisher1925theory,qmparis_quantum_2009,qmCaves_1994_Bures,new2liu2020quantum,qmpang_2014_metrology, qmpang_optimal_2017,qmhelstrom1976sld,qcrbfisher1925theory,qmparis_quantum_2009,qmCaves_1994_Bures,new2liu2020quantum}.
For diagonal entries, this reduces to
 \begin{eqnarray}
    \begin{aligned}
    J_{x_j}(T)=&4\langle \Delta^2S_{x_j}(T)\rangle,\\  
    \end{aligned}    
   \end{eqnarray}
where $\Delta^2S_{x_j}(T)=\langle S^2_{x_j}(T)\rangle-\langle S_{x_j}(T)\rangle^2$ is the variance of $S_{x_j}(T)$. Here all expectation values, $\langle \cdot\rangle$, are taken with respect to the initial probe state $|\psi_0\rangle$.

\paragraph{Optimal scheme}--- We focus initially on the minimal case of $N=2$ parameters and $d=2$ nodes. This captures all essential features of the problem, which we later generalize to an arbitrary number of parameters.  

The general Hamiltonian for this setting takes the form:
\begin{equation}
    H(t)=H_1(x_1,x_2,t)\otimes I + I \otimes H_2(x_1,x_2,t)+H_c(t),
\end{equation}
with $H_{1(2)}(x_1,x_2,t)=\Vec{f}^{1(2)}(\Vec{x},t)\cdot \Vec{\sigma}^{1(2)}$.
This model captures key scenarios in DQM, including two important cases.
The first scenario models independent, localized signals. Here, the parameters are encoded in distinct nodes via a Hamiltonian that decomposes as $H = H_1(x_1, t) \otimes I + I \otimes H_2(x_2, t)$, where each local Hamiltonian $H_{1(2)}$ depends only on its respective parameter $x_{1(2)}$. This corresponds to sensors interacting with separate local fields. The second scenario involves a global signal encoding all parameters, where the Hamiltonian terms $H_1$ and $H_2$ each depend on both $x_1$ and $x_2$. This models a single target interacting simultaneously with the entire sensor network, a configuration highly relevant for applications like quantum-enhanced radar.

Our objective is to identify the optimal scheme that achieves the highest effective QFI for estimating a global parameter $\theta_1$, defined as the linear combination
\begin{equation}
\theta_1 = w_{1}x_1 + w_{2}x_2.
\end{equation}
The effective QFI for $\theta_1$ is then given by (see the Supplemental Material~\cite{supp} for the derivation)
\begin{equation}
    J_{\theta_1}(T)=w_1^2J_{x_1}(T)+ w_2^2 J_{x_2}(T) +2w_1w_2 J_{x_1x_2}(T),\label{Jtheta}
\end{equation}
which can be compactly rewritten as
    \begin{align}
        J_{\theta_1}(T)
        &=4\langle \Delta^2S_{\theta_1}(T)\rangle.
    \end{align}
where 
$S_{\theta_1}(T)=w_1S_{x_1}(T)+w_2S_{x_2}(T)$. The problem of maximizing $J_{\theta_1}(T)$ is then reduced to the task of maximizing the variance of  $S_{\theta_1}(T)$. 

To identify the optimal strategies that maximize $\langle \Delta^2S_{\theta_1}(T)\rangle$, we write the generators $S_{x_1}(T)$ and $S_{x_2}(T)$ in integral form as~\cite{qmpang_optimal_2017}
\begin{align}\label{eq:S12}
     S_{x_1}(T)&=\int_0^T U^{\dagger}_{\rm tot}(t) \partial_{x_1} H(t) U_{\rm tot}(t)dt,\nonumber\\
      S_{x_2}(T)&=\int_0^T U^{\dagger}_{\rm tot}(t) \partial_{x_2} H(t) U_{\rm tot}(t)dt,
\end{align}
where $\partial_{x_{1(2)}} H(t)=\partial_{x_{1(2)}} H_1(t)\otimes I + I \otimes\partial_{x_{1(2)}} H_2(t)$.  
Combining these, the generator for $\theta_1$ can be written as
\begin{align}   
     S_{\theta_1}(T)=\int_0^T U^{\dagger}_{\rm tot}(t) \bm{V}_{\theta_1}(t) U_{\rm tot}(t)dt
\end{align}
where   
\begin{equation}
\bm{V}_{\theta_1}(t)=\bm{V}_{1}(t) \otimes I+ I \otimes \bm{V}_{2}(t),    \end{equation}
with
\begin{eqnarray}
\aligned
  \bm{V}_{1}(t)=w_1\partial_{x_{1}}H_{1}+w_2\partial_{x_2}H_{1}:=\vec{v}_1(t)\cdot \vec{\sigma},\\
  \bm{V}_{2}(t)=w_1\partial_{x_{1}}H_{2}+w_2\partial_{x_2}H_{2}:=\vec{v}_2(t)\cdot \vec{\sigma}.
  \endaligned
\end{eqnarray}

We now establish an upper bound for the variance of $S_{\theta_1}(T)$ as
\begin{equation}\label{QFI_theta}
    \langle \Delta S_{\theta_1}(T)^2\rangle\leq \left\{\int_0^T (\abs{\vec{v}_{1}(t)}+\abs{\vec{v}_{2}(t)} ) dt\right\}^2,
\end{equation}
where $|\vec{v}_{1}(t)|$ denotes the Euclidean length of $\vec{v}_{1}(t)$. The derivation of this bound proceeds in several steps. First, for any quantum observable, the variance in a state is bounded by the square of its spectral spread: $\langle \Delta S_{\theta_1}(T)\rangle\le\left(\lambda^{S_{\theta_1}}_{\max}(T)-\lambda^{S_{\theta_1}}_{\min}(T)\right)/2$, where $\lambda^{S_{\theta_1}}_{\max}(T)$ and $\lambda^{S_{\theta_1}}_{\min}(T)$ are the maximum and minimum eigenvalues of $S_{\theta_1}(T)$, respectively \cite{qmGiovannetti_2006_metrology}. Next, we bound these eigenvalues. Using the fact that $\lambda_{\max}(A+B)\leq \lambda_{\max}(A)+\lambda_{\max}(B)$ and $\lambda_{\min}(A+B)\geq \lambda_{\min}(A)+\lambda_{\min}(B)$ for Hermitian matrices \cite{weyl1949inequalities}, we have
\begin{eqnarray}
\aligned
\lambda^{S_{\theta_1}}_{\max}(T)\leq \int_0^T\lambda^{V_{\theta_1}}_{\max}(t) dt\leq \int_0^T (\abs{\vec{v}_{1}(t)}+\abs{\vec{v}_{2}(t)})dt,   \\
\lambda^{S_{\theta_1}}_{\min}(T)\geq \int_0^T\lambda^{V_{\theta_1}}_{\min}(t) dt\geq -\int_0^T (\abs{\vec{v}_{1}(t)}+\abs{\vec{v}_{2}(t)})dt.
\endaligned
\end{eqnarray}
From which we have 
\begin{align}
    \lambda^{S_{\theta_1}}_{\max}(T)-\lambda^{S_{\theta_1}}_{\min}(T)
    &\leq 2\int_0^T(\abs{\vec{v}_{1}(t)}+\abs{\vec{v}_{2}(t)})dt.
\end{align}
This directly leads to the upper bound for the effective QFI, 
\begin{equation}\label{eq:QFI}
    J_{\theta_1}(T)=4\langle\Delta S_{\theta_1}(T)^2\rangle\leq4\left\{\int_0^T(\abs{\vec{v}_{1}(t)}+\abs{\vec{v}_{2}(t)})dt\right\}^2.
\end{equation}

\begin{figure}[htbp]
\centering
\includegraphics[width=0.45\textwidth]{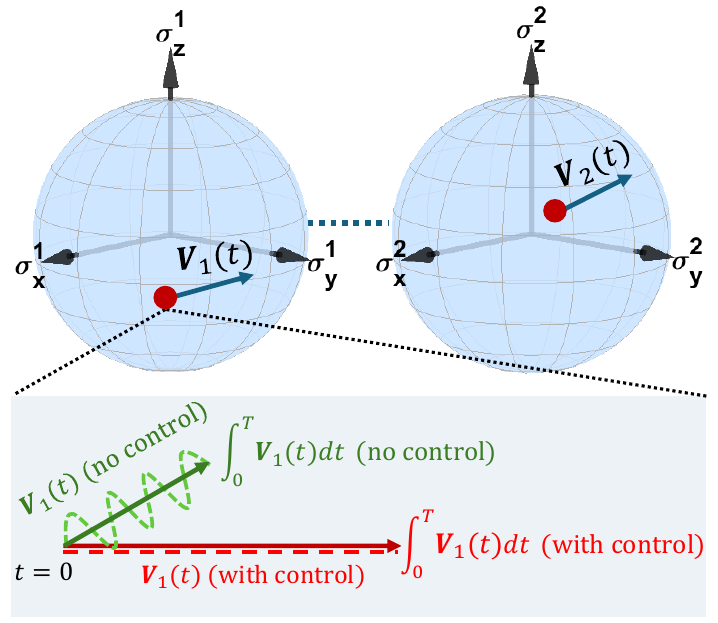}
\caption{A geometry perspective of QFI and generators with two nodes. Each node consists of one qubit, and the parameters are encoded through $\bm{V}_1(t)$ and $\bm{V}_2(t)$. 
Here we focus on node-1, provided with the optimal probe states, a larger displacement $\int_0^T \bm{V}_{1}dt$ indicates a higher QFI accumulated on this node. Here the green (red) dashed line represents the instantaneous velocity of $\bm{V}_{1}$ without (with) control, and the green (red) solid line represents the accumulated displacement without (with) optimal control.
}
\label{fig2}
\end{figure}

Crucially, this upper bound is saturable using only local control operations. This is achieved by implementing a control Hamiltonian with the separable form:
\begin{equation}
H_C(t)=H_{C1}(t)\otimes I+I\otimes H_{C2}(t).    
\end{equation}
 Under such local controls, the total evolution operator factorizes into a tensor product: $U_{\rm tot}(t)=U_1(t) \otimes U_2(t)$, where $U_i(t)$ is the evolution operator generated by the Hamiltonian of the $i$-th qubit plus its local control $H_{Ci}(t)$. This factorization allows the generator $S_{\theta_1}(T)$ to be decomposed into a sum of local terms
\begin{align}
    &S_{\theta_1}(T)=w_1S_{x_1}(T)+w_2S_{x_2}(T)\nonumber \\
    &=\int_0^T U^{\dagger}_{tot}(t)[\bm{V}_{1}(t)\otimes I+I\otimes \bm{V}_{2}(t)]U_{tot}dt\\
    &=\int_0^T [U^{\dagger}_1(t)\bm{V}_{1}(t)U_1(t)\otimes I+I\otimes U^{\dagger}_2(t)\bm{V}_{2}(t)U_2(t)]dt\nonumber,
\end{align}
where $\bm{V}_{{1(2)}}(t)=\vec{v}_{1(2)}\cdot \vec{\sigma}$. From a geometrical perspective, the vectors $\vec{v}_{1}(t)$ and $\vec{v}_{2}(t)$ represent instantaneous ``velocities'' in the generator's evolution. The local operations $U_{1(2)}(t)$ provide the ability to arbitrarily rotate the orientation of these vectors. To saturate the upper bound for the variance, the control strategy must align these effective velocities to constructively add their magnitudes. For instance, we can choose $U_1(t)$ and $U_2(t)$ to align each term along the $\sigma_z$ axis at every time point, 
\begin{eqnarray}
\aligned
U^{\dagger}_1(t)\bm{V}_{1}(t)U_1(t)&=\abs{\vec{v}_{1}(t)}\sigma_z,\\
U^{\dagger}_2(t)\bm{V}_{2}(t)U_2(t)&=\abs{\vec{v}_{2}(t)}\sigma_z. \label{oc}
\endaligned 
\end{eqnarray}
Under this strategy, the generator simplies to
\begin{align}
    S_{\theta_1}(T)=\left(\int_0^T \abs{\vec{v}_{1}(t)}dt\right) \sigma_z\otimes I+I\otimes \left(\int_0^T\abs{\vec{v}_{2}(t)}dt\right)\sigma_z.
\end{align}
whose eigenvalues satisfy 
\begin{equation}
\lambda^{S_{\theta_1}}_{\max}(T)-\lambda^{S_{\theta_1}}_{\min}=2\int_0^T (\abs{\vec{v}_{1}(t)}+\abs{\vec{v}_{2}(t)})dt.    
\end{equation}
This saturates the upper bound and achieves the maximal effective QFI. 

We note that the optimal control strategy is not unique: any set of local unitaries that consistently fixes the directions of $U^{\dagger}_1(t)\bm{V}_{1}(t)U_1(t)$ and $U^{\dagger}_2(t)\bm{V}_{2}(t)U_2(t)$ will saturate the bound. This flexibility allows for experimental adaptability.

Given the optimal total evolution operator $U_{\rm tot}^{\rm opt}(t) = U_1^{\rm opt}(t) \otimes U_2^{\rm opt}(t)$ that saturates the bound, we can derive the corresponding control Hamiltonian $H_C(t)=H_{C1}(t)\otimes I+I\otimes H_{C2}$. Let the time derivatives of the optimal local unitaries be governed by their respective Hamiltonians:
\begin{eqnarray}
    \aligned
    \dot{U}_1^{\rm opt}(t)&=-iH_1^{\rm opt}(t)U_1^{\rm opt},\\
    \dot{U}_2^{\rm opt}(t)&=-iH_2^{\rm opt}(t)U_2^{\rm opt},
\endaligned
\end{eqnarray}
we then have $\dot{U}_{\rm tot}^{\rm opt}(t)=-i[H_1^{\rm opt}(t)\otimes I+I\otimes H_2^{\rm opt}(t)]U_{\rm tot}^{\rm opt}(t)$. 

The total Hamiltonian is the sum of the free Hamiltonian $H_{\rm free}$ and the control Hamiltonian $H_C(t)$. To achieve the desired optimal evolution, this total Hamiltonian must generate the same dynamics as $H_1^{\rm opt}(t) \otimes I + I \otimes H_2^{\rm opt}(t)$. Therefore, the required control Hamiltonian is given by the difference: $H_{C1}(t)=H_1^{\rm opt}(t)-H_1(t)$ and $H_{C2}(t)=H_2^{\rm opt}(t)-H_2(t)$. We note that the optimal controls can depend on the parameters. In practice, the controls are implemented with the current best estimate of the parameters, which is then refined and updated adaptively throughout the estimation procedure.

For the special case of time-independent field sensing, where the system Hamiltonians $H_i(\vec{x},t)=\vec{f}^i(\vec{x})\cdot \vec{\sigma}$ are independent of time, the optimal control can be simplified. In this case, the operators  $\bm{V}_1=w_1 \partial_{x_1}H_1+w_2 \partial_{x_2}H_1$ and $\bm{V}_2=w_1 \partial_{x_1}H_2+w_2 \partial_{x_2}H_2$ have fixed directions. The optimal $U^{\rm opt}_{1}$ and $U^{\rm opt}_{2}$ can be taken as the identity operator with $H_1^{\rm opt}=H_2^{\rm opt}=\mathbf{0}$. The optimal control Hamiltonian reduces to $H_{C1}=-H_1$ and $H_{C2}=-H_2$, effectively canceling the free evolution. Again, this needs to be implemented adaptively as $H_{C1}=-H_1(\hat{\vec{x}})$ and $H_{C2}=-H_2(\hat{\vec{x}})$ with the estimated values of the parameters. 

Further simplification happens if the free Hamiltonian $H_1$ and $H_2$ are not only time-independent but also satisfy $[H_1,\bm{V}_1]=[H_2,\bm{V}_2]=0$ with $\bm{V}_1=w_1 \partial_{x_1}H_1+w_2 \partial_{x_2}H_1$ and $\bm{V}_2=w_1 \partial_{x_1}H_2+w_2 \partial_{x_2}H_2$. In this case, no control is needed since the free evolution operators, $U_1=e^{-iH_1t}$ and $U_2=e^{-iH_2t}$, preserve the directions of $\bm{V}_1$ and $\bm{V}_2$ with $U_1^\dagger \bm{V}_1 U_1=\bm{V}_1$ and $U_2^\dagger \bm{V}_2 U_2=\bm{V}_2$.  

In our control-enhanced scheme, implementing the optimal control Hamiltonian $H_C(\vec{x},t)$ requires knowing the exact values of $\vec{x}$. However, in reality, we do not know the exact values of $\vec{x}$, so the real control that we can apply to the node is $H_C(\hat{\vec{x}},t)$. In practice, rough estimators can first be obtained using non-optimal schemes; for example, one may employ separable probe states at each node to produce straightforward but suboptimal estimates of the parameters. 

This framework can be readily generalized to estimating \(N\) parameters with \(d\) nodes, where the \(k\)-th node contains \(n_k\) qubits. In this case, the free Hamiltonian becomes \(H_{\text{free}} = \sum_{k=1}^d \sum_{i=1}^{n_k} \vec{f}^{k}(\vec{x},t) \cdot \vec{\sigma}^{ki}\), with \(\vec{\sigma}^{ki}\) denoting the Pauli operators for the \(i\)-th qubit at the \(k\)-th node. In the supplementary material \cite{supp}, we show that the precision bound in this case is given by
\begin{equation}
\delta \hat{\theta}_1^2 \geq \frac{\vec{w}^T\vec{w}}{4 \left( \sum_{k=1}^d \sum_{i=1}^{n_k} \int_0^T |\vec{v}_{ki}(t)| \, dt \right)^2},
\label{generalNd}
\end{equation}
where \(\bm{V}_{ki}(t) = \sum_{j=1}^N w_j \partial_{x_j} H_{ki} := \vec{v}_{ki}(t) \cdot \vec{\sigma}\) with \(H_{ki} = \vec{f}^{k}(\vec{x},t) \cdot \vec{\sigma}^{ki}\). Notably, in this case the optimal control strategy can be implemented not only locally on each node, but also locally on each individual qubit within a node, further simplifying experimental requirements. We also show that the GHZ state is the optimal probe state, furthermore, the optimal measurement that achieves the bound can be implemented locally on each qubit.

\paragraph{Applications}---
To demonstrate the optimal scheme, we apply it to several practical applications including global clock synchronization, quantum radar, and frequency estimation.

We first consider the synchronization of two clocks, where the dynamics can be described as:
\begin{equation}
    H_{\rm free}=\Omega_1\sigma_z\otimes I+ I \otimes \Omega_2\sigma_z.
\end{equation}
and we are interested in estimating the difference of the two frequencies $\Omega_1-\Omega_2$. This corresponds to 
$\left(    \begin{array}{c}
         w_1  \\
         w_2 
    \end{array}\right)=\left(    \begin{array}{c}
        1  \\
         -1
    \end{array}\right)$. 
In this case, $H_1=\Omega_1\sigma_z$, $H_2=\Omega_2\sigma_z$, $\bm{V}_1=w_1\partial_{\Omega_1}H_1+w_2\partial_{\Omega_2}H_1=\sigma_z$, $\bm{V}_2=w_1\partial_{\Omega_1}H_2+w_2\partial_{\Omega_2}H_2=-\sigma_z$. These operators are all time-independent and they satisfy $[H_1, \bm{V}_1]=[H_2,\bm{V}_2]=0$. No controls are thus needed in this case. 
The maximal QFI can be obtained from Eq.(\ref{eq:QFI}) as 
\begin{equation}
     J_{\theta_1}(T)=4 \left\{\int_0^T\left(\abs{1}+\abs{-1}\right)dt\right\}^2=16T^2.
\end{equation}
This can be achieved by preparing the probe state as $\frac{|01\rangle-|10\rangle}{\sqrt{2}}$, which gives the highest precision $\delta \hat{\theta}_1^2 \geq \frac{1}{8T^2}$.
For comparison, an estimator for $\theta_1$ can also be obtained from a separable strategy, where $\Omega_1$ and $\Omega_2$ are individually estimated using a single qubit at each node and $\theta_1$ is subsequently obtained by computing their difference. In the separable strategy, the highest achievable precision is bounded by $\delta\hat{\theta}_1^2 \geq \frac{1}{4T^2}$. This is a factor of two worse than that achievable by the distributed strategy employing entangled states, consistent with previous results showing that global entanglement in the probe state provides an $O(d)$ precision enhancement in phase estimation \cite{qn2018multiparameter,qnge2018distributed,qnqian2019heisenberg}.

\begin{figure}[htbp]
\centering
\includegraphics[width=0.45\textwidth]{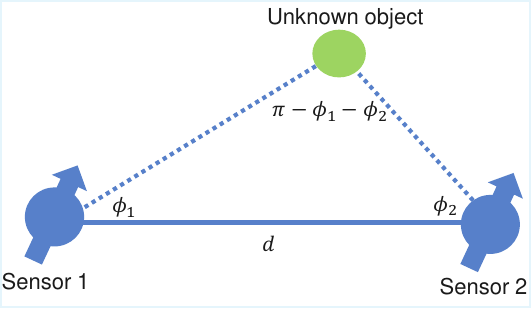}
\caption{A schematic illustration of the quantum radar principle in our setting. The goal is to determine the position of an unknown object by detecting the angle of its emitted signal at two spatially separated sensors. By adjusting the distance \( d \) between the sensors, the relative angles \( \phi_1 \) and \( \phi_2 \) can be used to infer the object’s location. Specifically, we extract the geometric information by measuring the angle \( \pi - \phi_1 - \phi_2 \), which is equivalent to estimating the linear combination \(\theta= \phi_1 + \phi_2 \).
}
\label{fig3}
\end{figure}

We next consider the estimation of a linear combination of $\phi_1$ and $\phi_2$ encoded in the Hamiltonian 
\begin{equation}
    H_{\rm free}=(\sin \phi_1 \sigma_x+\cos \phi_1 \sigma_z)\otimes I+ I \otimes (\sin \phi_2 \sigma_x+\cos \phi_2 \sigma_z).
\end{equation}
This can arise from detecting the position of an unknown object, as depicted in Fig.\ref{fig3}, where $\theta_1=\phi_1+\phi_2$ is the parameter of interest with $w_1=w_2=\frac{1}{{2}}$. In this case, we have $H_1=\sin \phi_1 \sigma_x+\cos \phi_1 \sigma_z$, $H_2=\sin \phi_2 \sigma_x+\cos \phi_2 \sigma_z$, $\bm{V}_1=w_1\partial_{\phi_1}H_1+w_2\partial_{\phi_2}H_1=(\cos \phi_1\sigma_x-\sin \phi_1 \sigma_z)$, $\bm{V}_2=w_1\partial_{\phi_1}H_2+w_2\partial_{\phi_2}H_2=(\cos \phi_2\sigma_x-\sin \phi_2 \sigma_z)$.
They are time-independent but $[H_{1(2)},\bm{V}_{1(2)}]\neq 0$. The optimal control can then be taken as $H_{C1}=-H_1$ and $H_{C2}=-H_2$. 
The maximal effective QFI for $\theta_1$ can be obtained from Eq.(\ref{eq:QFI}) as 
\begin{equation}
     J_{\theta_1}(T)=4 \left\{\int_0^T\left(\abs{1}+\abs{1}\right)dt\right\}^2=16T^2.
\end{equation}
This result can be compared to the maximum achievable QFI in the uncontrolled case, which is given by $J_{\theta_1}(T) = 16\sin^2(T)$ \cite{yuan2015optimal}. The controlled scheme's performance, scaling as $T^2$, stands in stark contrast to the oscillatory, bounded behavior of the uncontrolled case. This significant improvement highlights the critical role of control in enhancing distributed quantum metrology protocols.

As a final example, we consider two distributed time-dependent fields encoded in the Hamiltonian
\begin{equation}
    H_{\rm free}= \sin \Omega_1 t\,\sigma_z\otimes I+I\otimes \sin \Omega_2 t\,\sigma_z,
\end{equation}
where the parameter of interest is the sum of two frequencies $\theta_1=\Omega_1+\Omega_2$, corresponding to 
$w_1=w_2=1$. 
In this case, we have $H_1=\sin \Omega_1 t\sigma_z$, $H_2=\sin \Omega_2 t\sigma_z$, $\bm{V_1}(t)=-t\sin \Omega_1t \sigma_z$, $\bm{V_2}(t)=-t\sin \Omega_2t \sigma_z$.
The optimal control is always rotating $V_1(t)$ and $V_2(t)$ to a fixed direction, for example, along the direction of $-\sigma_z$. This can be achieved by applying $\pi$ pulses along the $x$-axis on the node 1 whenever $-t\sin\Omega_{1}t=0$ and $\pi$ pulses along the $x$-axis on the node 2 whenever $-t\sin\Omega_{2}t=0$. The maximal effective QFI is  
\begin{align}
      J_{\theta_1}(T)&=4 \left\{\int_0^T\left(\abs{-t\sin\Omega_1t}+\abs{-t\sin\Omega_2t}\right)dt\right\}^2\propto T^4,
\end{align}
which reaches the so-called ``Super-Heisenberg'' $O(T^4)$ \cite{qmnaghiloo_achieving_2017,mphou2021ac}. While for comparison, the maximum achievable QFI without controls is $J_{\theta_1}(T) = 2\left[\int_0^T({t\sin\Omega_1t}+{t\sin\Omega_2t})dt\right]^2$, a value significantly lower than its controlled counterpart.

\paragraph{Summary and outlook}---
In this work, we develop a comprehensive framework for designing optimal quantum sensing protocols to estimate a global parameter within a distributed sensor network. We derive a saturable upper bound for the effective QFI and provide systematic control strategies to achieve it.

Our analysis employs qubits as sensors at each node; a natural extension would be to generalize this to higher-dimensional quantum sensors. A further critical extension involves generalizing the framework to multi-parameter estimation, where the maximal QFI for an individual parameter no longer solely determines the overall estimation precision. The intricate interplay between all parameters must be accounted for \cite{wang2025tight,chen2024simultaneous,guhne2023colloquium,belliardo2021incompatibility}.

Another crucial direction for future work is to incorporate the effects of realistic noise. While we have established that the global GHZ state is the optimal probe in the noiseless setting considered here, the optimal probe choice is known to change dramatically under decoherence. For instance, many-body states such as Dicke states have demonstrated enhanced robustness against specific noise models \cite{pezze2018quantum,wang2002pairwise,kim1998influence,pezze2013ultrasensitive,saleem2024achieving}. Exploring their potential as noise-resilient probes in distributed sensing thus constitutes a highly promising avenue for future research.

\paragraph{Acknowledgments}--- Z.H. acknowledges Shilin Wang and Linmu Qiao for the helpful discussions. We acknowledge support from the ARO(W911NF-23-1-0077), ARO MURI (W911NF-21-1-0325), AFOSR MURI (FA9550-21-1-0209, FA9550-23-1-0338), DARPA (HR0011-24-9-0359, HR0011-24-9-0361), NSF (ERC-1941583, OMA-2137642, OSI-2326767, CCF-2312755, OSI-2426975), Packard Foundation (2020-71479), the Marshall and Arlene Bennett Family Research Program and the Quantised 2.0 and the Q-NEXT Center,  the Innovation Program for Quantum Science and Technology (2023ZD0300600), the Guangdong Provincial Quantum Science Strategic Initiative (GDZX2303007), the Research Grants Council of Hong Kong (14309223, 14309624, 14309022). This material is based upon work supported by the U.S. Department of Energy, Office of Science, National Quantum Information Science Research Centers and Advanced Scientific Computing Research (ASCR) program under contract number DE-AC02-06CH11357 as part of the InterQnet quantum networking project.

\framebox{\parbox{\linewidth}{
The submitted manuscript has been created by UChicago Argonne, LLC, Operator of 
Argonne National Laboratory (``Argonne''). Argonne, a U.S.\ Department of 
Energy Office of Science laboratory, is operated under Contract No.\ 
DE-AC02-06CH11357. 
The U.S.\ Government retains for itself, and others acting on its behalf, a 
paid-up nonexclusive, irrevocable worldwide license in said article to 
reproduce, prepare derivative works, distribute copies to the public, and 
perform publicly and display publicly, by or on behalf of the Government.  The 
Department of Energy will provide public access to these results of federally 
sponsored research in accordance with the DOE Public Access Plan. 
http://energy.gov/downloads/doe-public-access-plan.}}

\bibliography{main_text} 


\clearpage
\onecolumngrid

\begin{center}
    \mbox{\Large \textbf{Supplemental Material}}
\end{center}

\section{Optimal probe state}

Here we show that to estimate a linear combination of $N$ independent parameters as $\theta_1=\sum_{i=1}^N w_{i} x_i$, which are encoded within $d$ sensor nodes through $H_{\rm free}=\sum_{i=1}^d \Vec{f}^i(\Vec{x},t)\cdot \Vec{\sigma}^i
$, the optimal probe state can always be chosen as a $d$-qubits GHZ state.

We first begin by considering 2 unknown parameters encoded within 2 sensor nodes as
\begin{equation}
    H_{\rm free}(t)=H_1(x_1,x_2,t)\otimes I + I\otimes H_2(x_1,x_2,t),
\end{equation}
and we hope to find the optimal probe state that achieves the highest QFI for estimating a linear combination of unknown parameters as $\theta_1=w_{1} x_1+w_{2}x_2$. The QFI for $\theta_1$ is obtained as
\begin{align}
     J_{\theta_1}(T)&=\vec{w}^TJ_{\vec{x}}(T)\vec{w} \nonumber \\
     &= \begin{pmatrix} w_1 & w_2\end{pmatrix} \begin{pmatrix} J_{x_1}(T) & J_{x_{2}x_1}(T) \\ J_{x_{1}x_2}(T) & J_{x_2}(T) \end{pmatrix} \begin{pmatrix} w_1\\ w_2\end{pmatrix}  \nonumber \\
     &= w_1^2 J_{x_1}(T) + 2{w_1^2w_2^2} J_{x_{1}x_2}(T) + w_2^2 J_{x_2}(T)\label{Jtheta}
\end{align}
where $J_{x_1}(T), J_{x_2}(T)$ are the diagonal elements and $J_{x_{1}x_2}(T)$ is the off-diagonal element of the QFIM $J_{\vec{x}}(T)$. The matrix elements are given  as $J_{x_j}{(T)}=4\langle S^2_{x_j}(T)\rangle-4\langle S_{x_j}(T)\rangle^2$, $J_{x_1x_2}(T)=J_{x_2x_1}(T)=4{\rm Re}(\langle S_{x_1}(T)S_{x_2}(T)\rangle-\langle S_{x_1}(T)\rangle\langle S_{x_2}(T)\rangle)$ with $S_{x_j}(T)=i U_{\rm tot}^\dagger(T) \partial_{x_j} U_{\rm tot}(T)$ being the generators for $x_j$, $j=1,2$, and here $U_{\rm tot}(T)$ denotes the total unitary under control.

Our goal is to find the proper probe state that maximize 
\begin{equation}
    J_{\theta_1}(T)=w_1^2 J_{x_1}(T)+ w_2^2 J_{x_2}(T) +2w_1w_2 J_{x_1x_2}(T),\label{thetaQFI}
\end{equation}
We can rewrite Eq.\ref{thetaQFI} using generators as
\begin{align}
      J_{\theta_1}(T)&=4 \bra{\varphi_0}
  [w_1S_{x_1}(T)+w_2S_{x_2}(T)]^2\ket{\varphi_0}-4  \bra{\varphi_0} w_1S_{x_1}(T)+w_2S_{x_2}(T)\ket{\varphi_0}^2\nonumber \\
  &:= 4 \bra{\varphi_0}
  S_{\theta_1}(T)^2\ket{\varphi_0}-4  \bra{\varphi_0} S_{\theta_1}(T)\ket{\varphi_0}^2.
\end{align}
Now we show that the Bell state maximizes \begin{equation}
   J_{\theta_1}(T)= 4 \bra{\varphi_0}
  S_{\theta_1}(T)^2\ket{\varphi_0}-4  \bra{\varphi_0} S_{\theta_1}(T)\ket{\varphi_0}^2
\end{equation}
for any $S_{\theta_1}(T)$. Now we write $S_{\theta_1}(T)$ in the basis of Pauli matrices as 
\begin{equation}
    S_{\theta_1}(T):=\vec{s}_1(T)\cdot \vec{\sigma}\otimes I + I\otimes \vec{s}_2(T)\cdot \vec{\sigma},
\end{equation}
whose maximal and minimal eigenvalues are bounded by $\lambda_{\max}=\abs{\vec{s}_1(T)}+\abs{\vec{s}_2(T)}$ and $\lambda_{\min}=-\abs{\vec{s}_1(T)}-\abs{\vec{s}_2(T)}$  respectively. Thus 
\begin{equation}\label{eq:QFIBOUND}
    \begin{aligned}
         J_{\theta_1}(T)&\leq (\lambda_{\max}-\lambda_{\min})^2\\  &=4\left[\abs{\vec{s}_1(T)}+\abs{\vec{s}_2(T)}\right]^2 
    \end{aligned}      
\end{equation}
Now consider the Bell state \( \ket{\varphi_0} = \frac{1}{\sqrt{2}}(\ket{00} + \ket{11}) \) as the probe. Without loss of generality, we can assume \( \vec{s}_1(T) \) and \( \vec{s}_2(T) \) both point along the \( z \)-axis, i.e., \( \vec{s}_i(T) \propto \hat{z} \), since any other direction can be rotated into \( \hat{z} \) by a local unitary transformation to the node after the free evolution.
We have    \begin{equation}
       4[\bra{\varphi_0} S^2_{\theta_1}(T) \ket{\varphi_0}-\bra{\varphi_0}S_{\theta_1}(T) \ket{\varphi_0}^2]=4\left[\abs{\vec{s}_1(T)}+\abs{\vec{s}_2(T)}\right]^2
    \end{equation}
which saturates the bound in Eq.(\ref{eq:QFIBOUND}), indicating that the Bell state can serve as the optimal probe state for estimating two parameters encoded within 2 sensor nodes.

Next, we extend our results to estimating a global parameter $\theta_1$ as a linear combination of $N$ independent parameters encoded within $d$ sensor nodes as $\theta_1=\sum_{i=1}^N w_{i}x_i$. 
Similarly, the effective QFI for $\theta_1$ can be obtained as
\begin{equation}
J_{\theta_1}(T)
=\sum_{i=1}^{N} w_{i}^2\, J_{x_i}(T)
+ 2 \sum_{1 \leq i < j \leq N} w_{i} w_{j}\, J_{x_i x_j}(T).
\end{equation}
We can rewrite $J_{\theta_1}(T)$ using the generators of $S_{x_{i}}(T)$ as
\begin{align}
    J_{\theta_1}(T)&= \sum_{i=1}^d w_{i}^2 J_{x_i}(T) + 2\sum_{1\leq i<j\leq d} w_{i} w_{j} J_{x_i x_j}(T)\nonumber \\
    &=4 \langle (\sum_{i=1}^d w_{i}S_{x_i}(T))^2\rangle-4 ( \sum_{i=1}^d\langle  w_{i}S_{x_i}(T)\rangle)^2\\
    &:=4 \langle (S_{\theta_1}(T))^2\rangle-4 ( \langle S_{\theta_1}(T)\rangle)^2,
\end{align}
and we can rewrite $S_{\theta_1}(T)$ in the basis of Pauli matrices as 
\begin{equation}
    S_{\theta_1}(T):=\vec{s}_1(T)\cdot \vec{\sigma}\otimes I \otimes I  \otimes ...\otimes I + I\otimes \vec{s}_2(T)\cdot \vec{\sigma} \otimes I\otimes ...\otimes I+...+I\otimes I\otimes...\otimes \vec{s}_d(T)\cdot \vec{\sigma},
\end{equation}
whose maximal and minimal eigenvalues are bounded by $\lambda_{\max}=\sum_{i=1}^d\abs{\vec{s}_i(T)}$ and $\lambda_{\min}=-\sum_{i=1}^d\abs{\vec{s}_i(T)}$, respectively, so the maximal QFI is bounded as
\begin{equation}
    J_{\theta_1}(T)\leq4\left[\sum_{i=1}^d\abs{\vec{s}_i(T)}\right]^2.  
\end{equation}

Now consider the global GHZ state \( \ket{\varphi_0} = \frac{1}{\sqrt{2}}(\ket{00\ldots0} + \ket{11\ldots1}) \) as the probe. Similarly, under the assumption that each \( \vec{s}_i(T) \) points along the \( z \)-axis,
We have    \begin{equation}
       4[\bra{\varphi_0} S^2_{\theta_1}(T) \ket{\varphi_0}-\bra{\varphi_0}S_{\theta_1}(T) \ket{\varphi_0}^2]=4\left[\sum_{i=1}^d\abs{\vec{s}_i(T)}\right]^2,
    \end{equation}
indicating that the global GHZ state can be served as the optimal probe state in our settings. It is worth noting that the strategy presented here is just one approach to finding an optimal probe (with the aid of additional control). Other probe states may also maximize the variance of \( J_{\theta_1}(T) \) under specific conditions.

\section{Optimal control}
Now we show that provided with the optimal probe state, applying local control is sufficient to reach the highest effective QFI.

For estimating $N$ parameters encoded with $d$ nodes as $H_{\rm free}=\sum_{i=1}^d \Vec{f}^i(\Vec{x},t)\cdot \Vec{\sigma}^i$, with only local control, we have 
\begin{equation}
    U_{\rm tot}(t)=U_1(t)\otimes U_2(t)\otimes...\otimes U_d(t),
\end{equation}
and the generator $S_{\theta_1}(T)$ can be written as
\begin{align}
    S_{\theta_1}(T)&=\sum_{i=1}^N w_iS_{x_i}(T)\nonumber \\
    &=\int_0^T U^{\dagger}_{tot}(t)[\bm{V}_{1}(t)\otimes I\otimes  ... \otimes I+I\otimes \bm{V}_{2}(t)\otimes  ... \otimes I+ I \otimes ... \otimes I \otimes \bm{V}_{d}(t)]U_{tot}dt\\
    &=\int_0^T [U_{1}^{\dagger}(t)\bm{V}_{1}(t)U_{1}(t)\otimes I\otimes  ... \otimes I+I\otimes U_{2}^{\dagger}(t)\bm{V}_{2}(t)U_{2}(t)\otimes I \otimes ... \otimes I+ I \otimes ... \otimes I \otimes U_{d}^{\dagger}(t)\bm{V}_{d}(t)U_{d}(t)]dt\nonumber,
\end{align}
with 
\begin{equation}
    \bm{V}_{i}(t)=\sum_{j=1}^N w_j \partial_{x_j}H_i := \vec{v}_i(t)\cdot\vec{\sigma}.
\end{equation}
The optimal control aims to align $\bm{V}_{d}(t)$ to the same direction at different $t$ ($\sigma_z$ for example), and $U_{d}(t)$ satisfies the following requirement:
\begin{equation}
    U^{\dagger}_d(t)\bm{V}_{d}(t)U_d(t)=\abs{\vec{v}_d(t)}\sigma_z.
\end{equation}
Under the optimal control, we have
\begin{equation}
    S_{\theta_1}(T)=\left(\int_0^T \abs{\vec{v}_{1}(t)}dt\right) \sigma_z\otimes I\otimes ...\otimes I+I\otimes \left(\int_0^T\abs{\vec{v}_{2}(t)}dt\right)\sigma_z
    \otimes ...\otimes I+ I\otimes ... \otimes I \otimes \left(\int_0^T\abs{\vec{v}_{d}(t)}dt\right)\sigma_z,
\end{equation}
which leads to 
\begin{equation}
    \delta \hat{\theta}_1^2  \geq\frac{\vec{w}^T\vec{w}}{4\langle S^2_{\theta_1}(T)\rangle}=\frac{\vec{w}^T\vec{w}}{4\left(\sum_{i=1}^d \int_0^T\abs{\vec{v}_{i}(t)}dt\right)^2 }.\label{precision}
\end{equation}

This framework can be readily generalized to scenarios where each node contains multiple qubit sensors. For estimating \(N\) parameters with \(d\) nodes, where the \(k\)-th node contains \(n_k\) qubits, the free Hamiltonian becomes \(H_{\text{free}} = \sum_{k=1}^d \sum_{i=1}^{n_k} \vec{f}^{k}(\vec{x},t) \cdot \vec{\sigma}^{ki}\), with \(\vec{\sigma}^{ki}\) denoting the Pauli operators for the \(i\)-th qubit at the \(k\)-th node. This is a special case of the general Hamiltonian \(H_{\text{free}} = \sum_{k,i} \vec{f}^{ki}(\vec{x},t) \cdot \vec{\sigma}^{ki}\) where \(\vec{f}^{ki}(\vec{x},t) = \vec{f}^{k}(\vec{x},t)\). Since our analysis has no restriction on the functions of $\vec{f}^{ki}(\vec{x},t)$, it can be directly applied to obtain the precision bound:
\begin{equation}
\delta \hat{\theta}_1^2 \geq \frac{\vec{w}^T\vec{w}}{4 \left( \sum_{k=1}^d \sum_{i=1}^{n_k} \int_0^T |\vec{v}_{ki}(t)| \, dt \right)^2},
\label{generalNd}
\end{equation}
where \(\bm{V}_{ki}(t) = \sum_{j=1}^N w_j \partial_{x_j} H_{ki} := \vec{v}_{ki}(t) \cdot \vec{\sigma}\), and \(H_{ki} = \vec{f}^{k}(\vec{x},t) \cdot \vec{\sigma}^{ki}\). Notably, the optimal control strategy that achieves the maximal effective QFI can be implemented not only locally on each node, but also locally on each individual qubit within a node, further simplifying experimental requirements.

\section{Optimal measurement}
Now we show how to find the optimal measurement and make the QCRB achievable. From the Heisenberg uncertainty relation, we have
\begin{equation}
    \Delta O_{\theta_1} \Delta S_{\theta_1}(T) \geq \frac{1}{2} \abs{ \langle  [S_{\theta_1}(T), O_{\theta_1}]  \rangle }, \label{HUR}
\end{equation}
where $O_{\theta_1}$ is the observable. The optimal observable that saturates Eq.\ref{HUR} should satisfy
\begin{equation}
    (S_{\theta_1}(T)-\langle S_{\theta_1}(T)\rangle )\ket{\varphi_0}=i \gamma (O_{\theta_1}-\langle O_{\theta_1}\rangle )\ket{\varphi_0}, \gamma\in \mathbb{R} \label{om}
\end{equation}
 With the optimal control forcing $\bm{V}_{1}(t), \bm{V}_{2}(t),...,\bm{V}_{N}(t)$ to be proportional to $\sigma_z$, we can rewrite the generator as
\begin{equation}
    S_{\theta_1}(T)={s}_1(T)\,\sigma_z\otimes I \otimes I  \otimes ...\otimes I + I\otimes {s}_2(T)\, \sigma_z \otimes I\otimes ...\otimes I+...+I\otimes I\otimes...\otimes {s}_d(T)\cdot \sigma_z.
\end{equation}
With $\ket{\varphi_0}=(\ket{00...0}+\ket{11...1})/\sqrt{2}$ as the probe, $\langle S_{\theta_1}(T)\rangle=0$, we have
\begin{equation}
    (S_{\theta_1}(T)-\langle S_{\theta_1}(T)\rangle )\ket{\varphi_0} = \left(\sum_i^d s_i(T)\right) \frac{\ket{00...0}-\ket{11...1}}{\sqrt{2}},
\end{equation}
then we can choose the observable in the Heisenberg picture as $O_{\theta_1}=\sigma_x \otimes \sigma_x \otimes...\otimes\sigma_y$, which satisfies
\begin{equation}
    \langle O_{\theta_1}\rangle=0
\end{equation}
and
\begin{equation}
    O_{\theta_1}\ket{\varphi_0}=-i\frac{\ket{00...0}-\ket{11...1}}{\sqrt{2}},
\end{equation}
and we can choose $\gamma=-\sum_i^d s_i(T)$ to make Eq.\ref{om} satisfied. This corresponds to performing local projective measurements in the \( \sigma_x \) and \( \sigma_y \) basis on each qubit individually.

\end{document}